\documentclass{ieeeaccess}
\usepackage{cite}
\usepackage[utf8]{inputenc}
\usepackage{color}
\usepackage{amsfonts}
\usepackage{amsmath}
\usepackage{caption}
\usepackage{listings}
\usepackage{fancyvrb}
\usepackage{textcomp}
\usepackage{graphicx}
\usepackage[unicode=true,pdfusetitle,
 bookmarks=true,bookmarksnumbered=false,bookmarksopen=false,
 breaklinks=true,pdfborder={0 0 0},pdfborderstyle={},backref=false,colorlinks=true]
 {hyperref}
\usepackage{xcolor}

\definecolor{codegreen}{rgb}{0,0.6,0}
\definecolor{codegray}{rgb}{0.5,0.5,0.5}
\definecolor{codepurple}{rgb}{0.58,0,0.82}
\definecolor{backcolour}{rgb}{0.95,0.95,0.92}

\lstdefinestyle{mystyle}{
  backgroundcolor=\color{backcolour},   commentstyle=\color{codegreen},
  keywordstyle=\color{magenta},
  numberstyle=\tiny\color{codegray},
  stringstyle=\color{codepurple},
  basicstyle=\ttfamily\footnotesize,
  breakatwhitespace=false,         
  breaklines=true,                 
  captionpos=b,                    
  keepspaces=true,                 
  numbers=left,                    
  numbersep=5pt,                  
  showspaces=false,                
  showstringspaces=false,
  showtabs=false,                  
  tabsize=2
}
\begin{document}

\title{Benchmarking quantum co-processors in an application-centric, hardware-agnostic and scalable way}

\author{\uppercase{Simon Martiel}, 
\uppercase{Thomas Ayral},  
  \uppercase{Cyril Allouche}
}
\address{Atos Quantum Laboratory, Les Clayes-sous-Bois, France}

\doi{10.1109/TQE.2021.3090207}

\begin{abstract}
Existing protocols for benchmarking current quantum co-processors fail to meet the usual standards for assessing the performance of High-Performance-Computing platforms.
After a synthetic review of these protocols---whether at the gate, circuit or application level---we introduce a new benchmark, dubbed Atos Q-score\textsuperscript{TM}, that is application-centric, hardware-agnostic and scalable to quantum advantage processor sizes and beyond.
The Q-score measures the maximum number of qubits that can be used \emph{effectively} to solve the MaxCut combinatorial optimization problem with the Quantum Approximate Optimization Algorithm.
We give a robust definition of the notion of effective performance by introducing an improved approximation ratio based on the scaling of random and optimal algorithms. 
We illustrate the behavior of Q-score using perfect and noisy simulations of quantum processors.
Finally, we provide an open-source implementation of Q-score that makes it easy to compute the Q-score of any quantum hardware.
\end{abstract}


\maketitle

\section*{Introduction}

Recent years have witnessed great progress in the field of quantum technologies, whether on the hardware side---with growing computer sizes and quantum operation fidelities---or on the software side---with many algorithmic improvements.
This progress has, among other achievements, enabled recent claims that quantum advantage---the capacity for a quantum processor to outperform a classical machine---was attained by some of the most advanced Noisy, Intermediate Scale Quantum (NISQ, \cite{Preskill2018}) processors~\cite{Arute2019,Zhong2020}. 
However, these claims pertain to rather contrived, if not useless computational tasks carefully tailored for specific quantum processors.

In fact, the crucial milestone for the field to truly come of age is to identify hard, real-world computational problems whose solution can be accelerated by quantum computers. 
Many different hardware platforms with many different algorithmic ideas are vying for this goal today.
This diversity of quantum hardware and software candidates for quantum advantage requires a precise metric of success in order to appraise the relative power of each quantum computing stack for outperforming classical computers.
This metric will not only provide a much-needed synthetic overview of the current status of the field to end-users such as the high-performance-computing (HPC) community, but it will also help fuel the quantum community's efforts towards real-world applications.

This metric must fulfill a number of criteria to achieve these goals:
(i) \emph{Application-centric:} The metric must measure the ability to solve a hard, real-world computational problem that should be, at least to some extent, representative of a wide class of computational problems relevant to industry;
(ii) \emph{Hardware-agnostic:} The metric must not favor any hardware or software over another;
(iii) \emph{Scalable:} One must be able to compute the metric for large problem sizes. In particular, the scaling of the  classical processing time for computing the metric must be polynomial with the problem size.

In the field of classical HPC, these criteria are typically fulfilled by the LINPACK benchmark \cite{Dongarra2003} that is used to rank the TOP500 supercomputers.
In the field of quantum computing, a number of metrics have already been proposed in the literature. As we will explain in more detail in the next section, none of them fulfill all the above requirements. 

The purpose of this paper is to fill this gap by proposing a metric, dubbed "Q-score", that satisfies these requirements.
Essentially, the Q-score measures the maximum number of quantum bits that a quantum computer can use to solve a combinatorial optimization problem---the Max Cut problem---significantly better than a classical random algorithm.
In other words, it is an estimate of the largest (MaxCut) combinatorial optimization problem that can be solved better (compared to a random heuristic) on a quantum processor than on a classical computer. 
Q-score can be run and computed on any gate-based quantum hardware. 
Q-score is compatible with, although not restricted to, NISQ QPUs. In the version of Q-score presented in this article, we solve the combinatorial problem at hand, MaxCut, with a NISQ-compatible hybrid quantum-classical algorithm, the Quantum Approximate Optimization Algorithm (QAOA, \cite{Farhi2014}). Yet, any quantum algorithm tackling the MaxCut problem can in principle be considered as a suitable candidate. 
It takes into account the performance of the compilation.
An implementation is available under an open-source license.

To define the Q-score, we carefully investigate the size-dependence of the average performance of random and optimal classical algorithms, as well as the QAOA quantum algorithm, for solving the Max Cut problem on classes of random graphs.
The metric that we propose, akin to an improved approximation ratio, allows to measure non-trivial performance above the level of random classical algorithms.
Finally, we illustrate the behavior of the Q-score using noisy simulations with a depolarizing noise intensity compatible with today's NISQ processors.

This paper is organized as follows: we start by spelling out the desirable properties of quantum metrics and by reviewing the main existing quantum metrics (Section~\ref{sec:prior}). We then describe the Q-score protocol (Section~\ref{sec:protocol}) and discuss its properties (Section~\ref{sec:discussion}). We finally explain how to run this benchmark using an open-source script we provide online (Section~\ref{sec:doi}).

%%%%%%%%%%%%%%%%%%%%%%%%%%%%%%%%%%%%%%%%%%%%%%%%%%%%%%%%%%%%%%%%%%
\section{Characterizing quantum processors: goals and prior work}\label{sec:prior}

The careful design of Quantum Characterization, Verification and Validation (QCVV) protocols is crucial for assessing the potential of current and future quantum processing units (QPUs).
Several such protocols have been proposed in the recent years, with various levels of proximity to applications, scalability, fairness and practicality.

In this section, we start by laying out the QCVV criteria we deem to be most important from a High-Performance Computing (HPC) perspective. We then briefly review the main existing proposals and to what extent they fulfill these criteria.

\subsection{A High-Performance-Computing-driven list of criteria}
The first useful applications of quantum processors will likely be demonstrated in setups where quantum co-processors will be used as accelerators for performing very specific hard computational tasks within a  High-Performance-Computing (HPC) system.
The usefulness of the co-processor will be measured by comparing the performance of such a (possibly hybrid) computation with the performance of its purely classical counterpart.
With this in mind, we argue that useful QCVV protocols should fulfill the following three criteria:

(1) \emph{Application-centric}: The protocol should yield a single number (or a few) that unequivocally reflects the potential of a given QPU for solving a real-life HPC application. Ideally, the score of the QPU for this given application should be a proxy for how well the processor performs in general, i.e for other applications. This focus on applications and its "holistic" goal excludes protocols that narrow the characterization down to low-level components only, such as, e.g, gate quality or ability to sample specific classes of circuits (random or square circuits).

(2) \emph{Hardware-agnostic}: The protocol should put all the existing or future hardware technologies on an equal footing. 
In particular, it should not unduly favor a given technology over the others. Focusing on applications (see previous point) already ensures that the benchmark will incentivize hardware makers to make meaningful overall improvements, instead of focussed fine-tunings aimed at spoofing the benchmark as can more easily happen for gate- or circuit-level protocols. The application itself should also not be targeted to a given platform, and the difficulty of solving it should be representative of that of solving other hard problems (one wants to avoid niche applications that are contrived to perform well only in particular circumstances, and whose level of complexity is not easily comparable to other problems; here, we believe that our choice of QAOA, which is quite representative of variational algorithms, and of the MaxCut problem, whose encoding requires a reasonably low number of qubits, meets these demands---while being easily adjustable to better-performing future processors).   

(3) \emph{Scalable}: The protocol should be scalable to large numbers of qubits. In particular, the classical computational complexity for processing the quantum output and outputting the metric should be reasonably moderate. This constraint excludes protocols that involve classical computations that are exponentially costly in the number of qubits. 

\subsection{Prior proposals}
Most previously proposed QCVV protocols focus on gate-level and circuit-level characterization. We briefly review these protocols, which give valuable, albeit partial insights into the performance of a given QPU. We then turn to the previous attempts at characterizing QPUs from an application perspective.

\subsubsection{Gate-level protocols}\label{subsubsec:gatelevel}
In the past years, several protocols have been proposed to characterize the performance of the main low-level components of QPUs, quantum gates and sequences of gates, namely quantum circuits. 
The corresponding metrics give valuable information to compare different implementations of similar quantum technologies, such as two different experimental realizations of superconducting transmon processors. They also give indications about the ability of QPUs to run certain classes of quantum circuits.

The most widely used protocol for characterizing the gate-level quality of a QPU is \emph{Randomized Benchmarking} (RB) \cite{Magesan2011}. It yields the average fidelity $f$ or average error rate $\epsilon = 1 - f$ of a given gate set \cite{Proctor2017} while requiring only polynomial classical resources (provided potentially exponential compilation overheads are avoided, such as in Direct Randomized Benchmarking \cite{Proctor2019a}). It is also robust to state preparation and measurement (SPAM) errors. These two aspects are major advantages over direct fidelity estimation protocols.
On the flip side, RB is not application-centric: it gives little information as to the performance of circuits, let alone applications. Indeed, structured circuits (as opposed to the random circuits used in RB) are more sensitive to errors than randomized circuits, and thus one can hardly predict the performance of a structured circuit given the RB metrics of its gate set (see \cite{Proctor2020} for protocols that use structured circuits). One major reason for this deficiency is that RB gives little information about crosstalk errors, which influence the performance of a QPU at the circuit level (although we note that recent works propose ways of extending RB to crosstalk estimation \cite{McKay2020}).

Another widely used protocol that goes beyond the measurement of the mere average fidelity of a gate set is \emph{Gateset Tomography} (GST) \cite{Blume-kohout2013,Merkel2013,Greenbaum2015}. This quantum process tomography method yields the specific noise model of each quantum operation that a QPU is able to perform, including gates, state preparation, and measurement. Once the so-called GST gauge has been properly fixed (see, e.g, \cite{DiMatteo2020}), average fidelities can be extracted from the noise models. More interestingly, the noise models can be used as inputs to circuit-level simulations. These simulations are generically exponentially costly in the number of qubits, but can yield precise information about the behavior of a given circuit executed on a given processor.
However, if GST is realized at the one-qubit and two-qubit level only, crosstalk effects beyond two-qubit crosstalk, which are suspected to play an important role in NISQ devices, will be neglected.
Going beyond one-qubit and two-qubit errors to capture those effects is possible, but requires a cost in terms of classical processing \emph{and} amount of data to be collected from the QPU that scales exponentially with the number of qubits. Thus, GST is hardly scalable for real-world applications.

Recently, a protocol called \emph{Cycle Benchmarking} (CB) \cite{Erhard2019} has been proposed to go beyond the limitation of RB and GST to the characterization of few-qubit error processes. While RB and GST require a number of experiments that scales exponentially with the number of qubits involved in the quantum process to be characterized (whether an error or a gate), thus limiting them to very few qubits, CB allows to characterize processes acting on much larger registers. This applies to crosstalk errors (see previous paragraph), but also to multi-qubit gates like the M{\o}lmer-Sorensen gate of trapped-ion processors that act on multiple (even all) qubits in a register. 
In additional, CB is robust to SPAM errors \cite{Erhard2019}. However, as a gate-level protocol, it cannot be used to characterize the potential of a QPU at an application level.

%QPUs with operations acting only a few qubits. Indeed, beyond the crosstalk characterization issue we raised in the previous paragraph, some technologies such as trapped-ion QPUs provide operations like the M{\o}lmer-Sorensen gate that act on multiple (even all) qubits in a register. CB can detect crosstalk and is robust to SPAM errors \cite{Erhard2019}. However, as a gate-level protocol, it cannot be used to characterize the potential of a QPU at an application level. 

\subsubsection{Circuit-level protocols}
A number of protocols has been proposed to measure the ability of QPUs to run certain classes of circuits.

One such protocol is the \emph{Quantum Volume} (QV) \cite{Cross2019} metric, and its generalization to non-square circuits, \emph{Volumetric Benchmarks} (VB) \cite{Blume-Kohout2019}.
They measure the ability of a QPU to prepare a random state given a certain number of qubits (circuit width) and a certain gate count (circuit depth). While QV looks only at square circuits (with equal width and depth), which fails to capture algorithms that do not involve square circuits (like Shor's algorithm), VB lifts this limitation. However, the core metric of QV/VB, namely the heavy output generation probability (HOV, \cite{Aaronson2016}), requires the exponentially costly computation of probability amplitudes (to compute the set of heavy outputs). These approaches are thus not scalable. Furthermore, they focus on classes of random circuits, making them hardly suitable for assessing the performance of a QPU on a real application.

Related protocols, dubbed \emph{Cross-Entropy Benchmarking} (XEB) and \emph{Cross-Entropy Fidelity} \cite{Arute2019,Neill2017}, have been recently proposed and used to compare the ability of a QPU to generate random states with that of a classical computer. Like QV and VB, these protocols require classical resources that are exponential in the number of qubits, thereby limiting their scalability. Second, the task they seek to optimize is the sampling of bitstrings measured after executing families of random circuits. The performance of a given QPU in solving such a specific problem hardly qualifies as application-centric in the absence of a straightforward extrapolation of the corresponding metric to real-world applications.

Recently, Ref.~\cite{Mills2020} proposed a series of benchmarks that comprise the QV, VB and XEB metrics together with the $l_1$ norm to compare probability distributions. Similar limitations in terms of the classical complexity to compute the metric and difficulty to use it as a proxy for an actual application also apply to this work.

Ref. \cite{Proctor2020} recently proposed a protocol based on the "mirroring" concept (also used in RB) that allows to get rid of the exponential classical effort that plagues the previous circuit-level protocols. Yet, the ability to use this other circuit-level metric to reliably predict the behavior of a given QPU for a real application remains to be investigated.

%%%%%%%%%%%%%%%%%%%%%%%%%%%%%%%%%%%%%%%%%%%%%%
\subsubsection{Application-level protocols}
%%%%%%%%%%%%%%%%%%%%%%%%%%%%%%%%%%%%%%%%%%%%%%

We now turn to application-level protocols.

One of the most promising applications of quantum processors is the field of quantum many-body physics, since quantum processors are by construction quantum many-body systems with a large number of quantum bits interacting with one another in a controlled fashion.

Ref.~\cite{Dallaire-Demers2020} recently proposed a metric dubbed \emph{Fermionic Depth} (FD) to quantify the ability of a QPU to tackle a quantum many-body problem. The prototypical many-body problem chosen in this work is the one-dimensional Fermi-Hubbard model, whose ground-state energy in the infinite-size limit, $E_{\infty}^{\mathrm{exact}}$, can be computed exactly in polynomial time on a classical computer via the so-called Bethe ansatz method \cite{Lieb1968}.  
The protocol consists in computing, with a QPU, the approximate ground-state energy of this model $E_{L}$ for different (linear) sizes $L$, and then returning the deviation
to the exact energy at infinite size, $\Delta E_{L}=E_{L}-E_{\infty}^{\mathrm{exact}}$. 
In practice, due to the limited coherence of current (NISQ) processors, $E_L$ is computed via a hybrid quantum-classical method, the Variational Quantum Eigensolver (VQE, \cite{Peruzzo2014}) method, as opposed to fully coherent algorithms like the Quantum Phase Estimation algorithm, that are not suitable for non-error-corrected QPUs.
Due to decoherence effects, the corresponding $\Delta E_{L}$ curve is going to display a minimum at a given size $L^*$, dubbed the \emph{fermionic length} of the QPU under investigation. This fermionic length thus gives an indication about the maximum size of a fermionic problem that a given QPU can handle.

The predictive power of this metric for problems outside the 1D Fermi-Hubbard model remains to be investigated: whether the fermionic length estimated for a one-dimensional problem is related to the fermionic length that can be achieved for two-dimensional quantum many-body problems is an open question. Indeed, those two-dimensional problems, which are among the hardest to tackle with the most advanced classical algorithms, display phenomena (high-temperature superconductivity, pseudogap phase, ...) that are radically different from one-dimensional problems.

Quantum chemistry problems, on the other hand, usually feature interactions between many orbitals, whereas the Hubbard model has only local interactions, raising the question of the relevance of the fermionic length for chemistry problems. We note that Ref.~\cite{McCaskey2019} proposed a chemistry-based benchmark of quantum processors, albeit with a focus on small molecules only and therefore no clear path towards scalability yet.

Finally, Ref.~\cite{Dong2020} proposed an extension of the LINPACK benchmark (that is used to rank classical supercomputers) to a quantum setting. The protocol consists in solving a linear system of equations $A x = b$, with $A$ a random dense matrix, by outputting an approximate solution $g(A) |b\rangle $ with $g(x)$ a polynomial approximation of $x^{-1}$.
While this protocol avoids the usual read-in problem (it does not require the use of a QRAM to load $A$ from classical data) through a block-encoding method (random circuits $U_A$ are used such that one of the blocks of this unitary is $A$, with $A$ a random dense matrix), its measure of success consists in comparing the output vector $g(A) |b\rangle $ to the actual solution (in addition to a measure of the wall-clock time). This entails an exponential classical cost (through e.g a cross-entropy test), which limits the scalability of the method.

\section{The protocol}\label{sec:protocol}
In this section we describe our benchmark metric proposal. Similarly to other benchmark proposals, Q-score works by iteratively testing a quantum co-processor using a scalable test $T_n$ indexed by a \emph{problem size} $n$. Naturally, the score will be the largest problem size $n^\star$ such that $T_{n^\star}$ holds.

Informally the test consists in:
\begin{itemize}
    \item[(a)]Picking a collection of random graphs of size $n$
    \item[(b)]Running a QAOA-MaxCut algorithm on these graphs and computing $C(n)$, the average of the expected cut cost for each instance
    \item[(c)] Computing a score $\beta(n)$ that depends on $C(n)$ and testing $T_n: \beta(n) > \beta^\star$ for some constant $\beta^\star$.
\end{itemize}

The next subsection is dedicated to the description of this test $T_n$. The detailed explanation of the various choices described in this section can be found in section \ref{sec:discussion}.

\subsection{Description of the test}
Our test $T_n$ consists in running a Quantum Approximate Optimization Algorithm ({\bf QAOA}) for a MaxCut instance of size $n$. We now describe the settings in which the algorithm is run, and how its performance is assessed for a given instance size.

\paragraph{ The circuit implementation.} We assume that we tackle instances using the standard QAOA Ansatz as described in \cite{Farhi2014}. Given a graph $G=(V,E)$ (with $V$ and $E$ the vertex and edge set, respectively) and a depth parameter $p$, we implement the parameterized circuit:
\begin{equation}
    U(\boldsymbol{\gamma}, \boldsymbol{\beta}) = \prod_{1 \leq q\leq p} e^{-i\frac{\beta_q}{2} H_0}e^{-i\frac{\gamma_q}{2} H_G}
    \label{eq:U_def}
\end{equation}
where $H_0 = -\sum_{1\leq i \leq n} \sigma_x^{(i)}$ and 
\begin{equation}
    H_G = \sum_{i, j \in E} \sigma_z^{(i)}\sigma_z^{(j)} - \frac{|E|}{2}. 
    \label{eq:HG_def}
\end{equation}
Here, $\sigma_x$ and $\sigma_z$ denote the Pauli X and Z operators, and $|E|$ is the number of edges in the graph.

In practice, each rotation $e^{-i\frac{\gamma_q}{2}\sigma_z^{(i)} \sigma_z^{(j)}}$ is decomposed using a sub-circuit of $2$ CNOT gates and a single $R_Z$ rotation.
The propagator $e^{-i\frac{\beta_q}{2} H_0}$ is implemented using a wall of $R_X$ gates.

\paragraph{The classical optimizer.} The classical optimization routine used to minimize the Ansatz energy is COBYLA \cite{Powell1994}. This optimizer behaves well in perfect settings and for shallow circuits (i.e circuits with a low number of parameters). Since we expect that increasing the depth $p$ of the Ansatz will probably only degrade performances, this choice seems reasonable. Indeed, although increasing the depth leads to a larger variational search space and thus a potentially lower variational energy, on NISQ QPUs, larger depths also lead to an increased sensitivity to noise and thus usually degraded performances. This is what we observe in Fig.~\ref{fig:noisy_simulations_grid} except for noise levels that are very low compared to the noise levels reported for NISQ processors (this phenomenon was also observed in \cite{Arute2020b}). COBYLA is thus a sensible choice for current QPUs.

\paragraph{Computing the score.}
For a given size $n$, we run a QAOA-MaxCut on $100$ random graphs in $\mathcal{G}(n, p=\frac{1}{2})$, the distribution of Erd\"os-Renyi graphs obtained by taking an empty graph and connecting each pair of vertices with probability $\frac{1}{2}$. These graphs are relatively dense and constitute a standard class used for benchmarks.
Given $C(n)$, the average of the energies (multiplied by $-1$) produced by QAOA over these $100$ graphs, we compute the following ratio:
\begin{equation}
    \beta(n) = \frac{C(n) - \frac{n^2}{8}}{\lambda n^{3/2}}.
    \label{eq:qscore_def}
\end{equation}
We say that the quantum processor passes the test for this size $n$ if $\beta(n) > \beta^\star$.
Here, the threshold $\beta^\star \in ]0, 1[$ dictates how demanding is the test: a test with $\beta^\star = 0$ can be passed by a simple coin toss, while a test with $\beta^\star = 1$ can only be passed by an exact solver. Hence $\beta^\star$ can be seen as fraction of performance between a naive randomized algorithm and an exact solver. In practice, the threshold $\beta^\star$ is arbitrarily set to $0.2$. We take $\lambda = 0.178$ (see discussion below, section~\ref{sec:discussion}). We also fix the number of shots (repetitions) to be used to get the estimate of the QAOA energy for a given graph to 2048 per $(\boldsymbol{\beta},\boldsymbol{\gamma})$.

The final Q-score is the largest $n$ such that this test succeeds, i.e
\begin{equation}
    n^\star \equiv \max \lbrace n\in \mathbb{N}, \beta(n) > \beta^\star  \rbrace.
\end{equation}

\paragraph{Remarks.} The choice of $\beta^\star$ is somewhat arbitrary. $\beta^\star$ was set so that a QAOA of depth $p=1$ running on a perfect quantum processor will pass the test and will have an infinite Q-score. (As will be seen later [Fig.~\ref{fig:maxcut_scaling}], for $p=1$, $\beta^\mathrm{Q}(n) \approx 40\%$ for a perfect QPU).
In practice, the Q-score implementation we provide is parameterized by this $\beta^\star$.
Moreover, it is usually not necessary to iteratively try each instance size until the test fails, since $\beta(n)$ is expected to be a monotonically decreasing function of $n$.
This implies that one can employ a dichotomic search in order to find $n^\star$, the largest $n$ such that $\beta(n^\star) > \beta^\star$. Our implementation supports both iterative evaluation and dichotomic search.

\subsection{Illustration: perfect and noisy simulations}\label{subsec:illustration}
\begin{figure}
    \centering
    \includegraphics[width=1.0 \columnwidth]{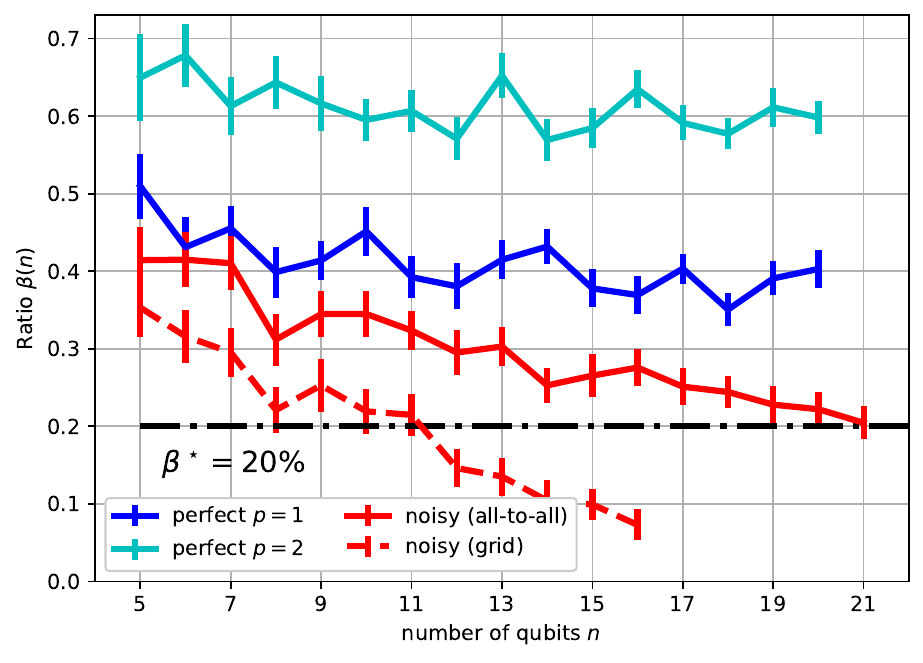}
    \caption{Evolution of $\beta(n)$ for different simulated QPUs: perfect QPU with $p=1$ (blue), $p=2$ (cyan), and noisy QPU with a depolarizing noise model (see text), with $p=1$, all-to-call connectivity (solid red lines), and grid connectivity (dashed red lines). The dash-dotted black line shows the 20\% threshold above which the Q-score test is passed. The error bar is the standard error of the mean score over 100 graphs.}
    \label{fig:noisy_simulations}
\end{figure}

To illustrate the meaning of the Q-score, we simulated the behavior of QAOA-MaxCut on various Quantum Processing Units (QPUs) using the Atos Quantum Learning Machine (QLM).

We started by running QAOA-MaxCut on a perfect (noiseless) QPU for two values of the number $p$ of QAOA layers.
As expected, we see, in Figure~\ref{fig:noisy_simulations}, that the score increases with an increasing $p$ due to an increased expressivity of the QAOA ansatz.
We also observe that the ratio $\beta(n)$ achieved by this perfect QPU is roughly constant as $n$ increases, with $\beta(n) \approx 40\%$ for $p=1$, and $\beta(n)\approx 60\%$ for $p=2$.
This means that QAOA executed on a perfect quantum processor achieves scalings within 40 \% (resp. 60 \%) of the optimal scaling $\lambda n^{3/2}$ (after subtraction of the leading $n^2/8$ term).

We compare this behavior to the scores obtained with simulations of noisy QPUs. 
We choose a simple depolarizing noise model with a level of noise that is consistent with today's NISQ processors. 
More specifically, we add depolarizing noise after each gate, with an average error rate of $$\epsilon_2=2\%$$ for two-qubit gates (in comparison, the two-qubit error rates reported for IBM Johannesburg~\cite{IBMQX}, Google Sycamore~\cite[Fig.2, Table II]{Arute2019}, Rigetti Aspen 7~\cite{RigettiWebSite} and ionQ~\cite{Wright2019}, are, respectively, 0.2\%, 0.62\%, 4.8\% and 2.5\%), and $$\epsilon_1=0.4\%$$ for one-qubit gates (this factor of 5 between the one- and two-qubit error rates is observed in typical superconducting and trapped-ion architectures, with reported one-qubit error rates of 0.041\%, 0.16\%, 0.77\% and 0.5\% for the four aforementioned platforms).
For the sake of simplicity, we assume perfect initialization and readout, and neglect noise during idling periods.

We observe that the ratio $\beta(n)$ achieved with a noisy QPU is, as expected, lower than with a perfect QPU. More importantly, it decreases with the problem size $n$ (i.e the number of qubits): larger problems require longer circuits and hence lead to an increased sensitivity to noise.
Moreover, a limited connectivity (\emph{e.g} a grid connectivity) leads to a decreased ratio, since these connectivity constraints require the original QAOA circuit to be optimized to comply with the constraints. This optimization, carried out following a method described in \cite{Hirata2009, martiel2020architecture} using one of Atos QLM's compilation plugins, leads to longer circuits and hence degraded performance in the presence of noise.

From these simulations, we can infer that the Q-score for a noisy QPU with a grid connectivity is $n^\star = 11$. For the noisy QPU with an all-to-all connectivity, we can infer that $n^\star = 21$. For perfect QPUs, QAOA achieves an infinite Q-score. 

\begin{figure}
    \centering
    \includegraphics[width=1.0 \columnwidth]{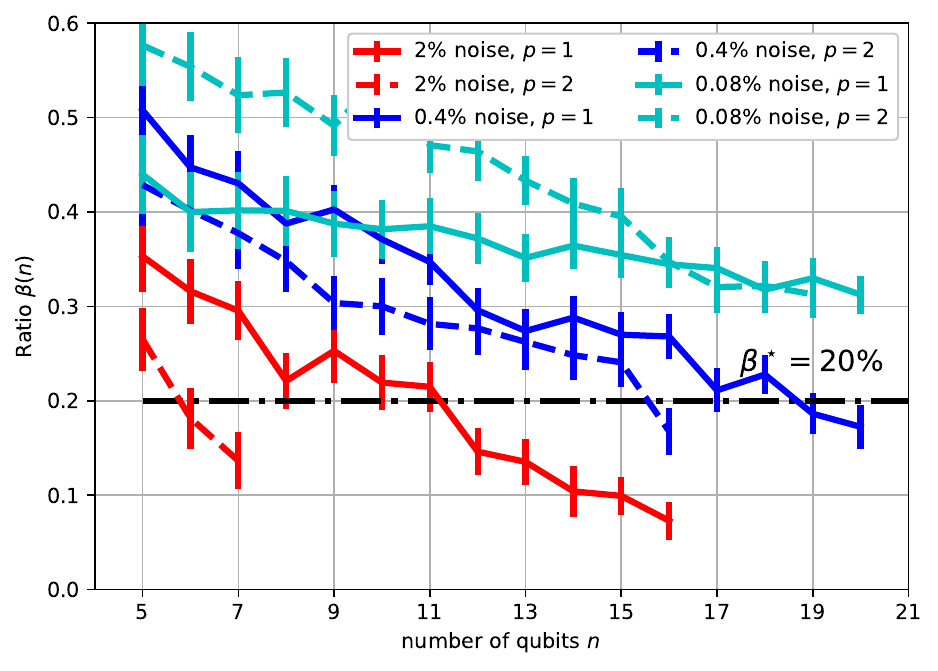}
    \caption{Evolution of $\beta(n)$ for different levels of depolarizing noise and number of QAOA layers $p$, with grid connectivity: $p=1$ (solid lines), $p=2$ (dashed lines). The dash-dotted black line shows the 20\% threshold above which the Q-score test is passed. The error bar is the standard error of the mean score over 100 graphs.}
    \label{fig:noisy_simulations_grid}
\end{figure}
In Fig.~\ref{fig:noisy_simulations_grid}, we exemplify the tradeoff between increasing the expressivity of QAOA's ansatz (by increasing the number of layers $p$) and curbing the impact of noise. As expected, we observe that for the higher noise levels ($\epsilon_2=2\%$  and 0.4 \%), a larger $p$ leads to a decreased Q-score (for $\epsilon_2=2\%$, the Q-score is 5 for $p=2$ while it is 11 for $p=1$) because the detrimental effect of noise outweighs the expressivity gain. At the lowest noise level ($\epsilon_2=0.08 \%$), the situation is more constrasted: while, for small graph sizes, a larger $p$ leads to a larger $\beta(n)$ (as is the case for the noiseless case, as shown in Fig.~\ref{fig:noisy_simulations}), for larger graph sizes (or numbers of qubits), noise penalizes longer circuits ($p=2$) over shorter ones ($p=1$), counterbalancing the increase in representativity due to a larger $p$.

Let us stress that this example also shows that beyond assessing the quality of the hardware for solving QAOA-MaxCut, the Q-score also assesses the performance of the software stack: for instance, a better compiler to optimize for connectivity constraints will lead to an increased $\beta(n)$ and hence to an increased Q-score. 
This is a major advantage of Q-score over lower-level metrics as improving \emph{both} the software stack and the hardware is crucial in the overall advancement of the field, whether at the algorithmic level, at the compilation stage, or via noise-mitigation techniques. While the risk of some users deliberately fine-tuning their software to spoof the Q-score benchmark exists, we believe that it is outweighed by the overall benefit that the community will draw from a healthy competition to increase Q-score.

\section{Discussion}\label{sec:discussion}
In this section, we discuss the various choices made in this proposal. First of all, let us recall briefly what we need to achieve.
\subsection{The algorithm choice}
We are not looking at finding a discerning metric for quantum supremacy. Our goal is simply to consider an application that is both representative of practical needs from the industry and challenging for current hardware platforms.

\paragraph{The choice of QAOA-MaxCut} Most, if not all, proposed algorithms compatible with the NISQ era, are variational algorithms. It thus seems natural, in an application-centric benchmark, to focus on this type of algorithms. Among all these propositions, we need one that fits a particular set of requirements.
First, the algorithm should be scalable, in the sense that one should be able to rather smoothly increase the problem size in order to isolate the precise threshold were the quantum co-processor fails. Combinatorial optimization problems usually fit this criterion quite easily.
Moreover, we also need the test to be efficiently computable. By averaging over a simple class of random instances, we can deduce asymptotic values for usually intractable quantities (see next subsection). This might be hard to do efficiently for other classes of problems. Hence, the Quantum Approximate Optimization Algorithm seems to be a good candidate that fits these needs. 
We chose the MaxCut problem for the simple reason that it is both simple to implement and simple to analyze. For instance, it is possible to know the average number of entangling gates required in the Ansatz, even after compilation and optimization. This would not be the case were we to consider problems that involved clauses over more that $2$ variables (mainly due to the variability of the literature in architecture-aware phase polynomial synthesis algorithms \cite{Nash_2020, martiel2020architecture,degriend2020architectureaware} or other less competitive SWAP-based routing techniques).

\paragraph{The choice of the class $\mathcal{G}(n, \frac{1}{2})$} This class of graphs is quite standard in random graph literature and has a predictable behavior with regard to the MaxCut problem. Moreover, they constitute a class of dense graphs, with half of their possible edges present (on average).
QAOA-MaxCut are often run using $k$-regular graphs for the simple reason that these graphs are very sparse. In fact their edge density decrease with their size. We suggest that most real-world applications will not have this property. Hence the choice of $\mathcal{G}(n, \frac{1}{2})$. One could relax a bit the test by picking a class $\mathcal{G}(n, f(n))$ with $f(n) = o(\frac{1}{n})$, that is a class of graph where edges are picked uniformly with a probability that decreases with $n$, but such that the average number of edges $f(n)\frac{n^2}{2}$ still grows faster that $n$.
Another potential choice would be to consider bipartite graphs. These graphs can be perfectly cut and thus hard instances for QAOA. They are however trivial to deal with classically, and as such, do not constitute an interesting benchmark target.

\subsection{Test definition and approximation ratio} \label{subsec:ratio}
In this subsection, we detail the 
%thought process
reasoning
behind the definition of the score (Eq.~(\ref{eq:qscore_def})) and the corresponding success criterion.

\paragraph{The usual approximation ratio and its lower bound.}
We recall that the algorithm is run on Erd\"os-Renyi graphs $G$ of fixed size $n$ and with edge probability $\frac{1}{2}$, denoted $\mathcal{G}\left (n, \frac{1}{2}\right)$.
A standard way to evaluate the performance of an approximation heuristic such as QAOA is to consider the \emph{approximation ratio} $\alpha(G) = \frac{C(G)}{C_\mathrm{max}(G)}$, where $C(G)$ is the score of the worst solution that can be produced by the heuristic and $C_\mathrm{max}(G)$ is the cost of the optimal solution for the given graph $G$. Since we are dealing with a randomized algorithm, this quantity translates into $\alpha^\mathrm{Q}(G) = \frac{\mathbb{E}^\mathrm{Q}[C(G)]}{C_\mathrm{max}}$ where $\mathbb{E}^\mathrm{Q}[C]$ would be the expected score of a solution produced by QAOA. (With an infinite number of shots, $\mathbb{E}^\mathrm{Q}[C(G)] = -\langle \Psi(\boldsymbol{\gamma},\boldsymbol{\beta}) | H_G | \Psi(\boldsymbol{\gamma},\boldsymbol{\beta}) \rangle$, with $|\Psi(\boldsymbol{\gamma},\boldsymbol{\beta})\rangle = U(\boldsymbol{\gamma},\boldsymbol{\beta}) |0\rangle ^{\otimes n}$, see Eqs.~(\ref{eq:U_def}) and (\ref{eq:HG_def})).

\begin{figure}
    \centering
         \includegraphics[width=1.0\columnwidth]{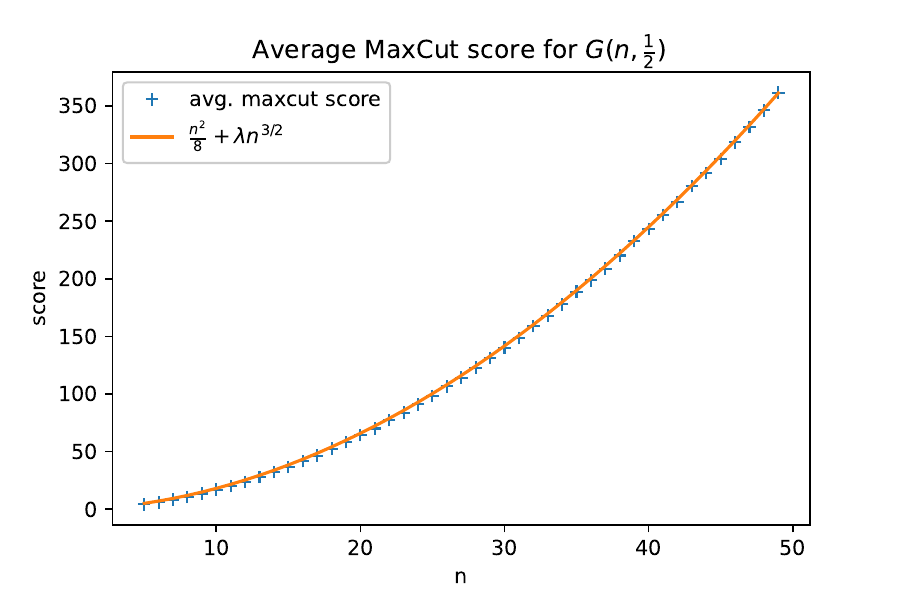}\\%
         \includegraphics[width=0.9\columnwidth]{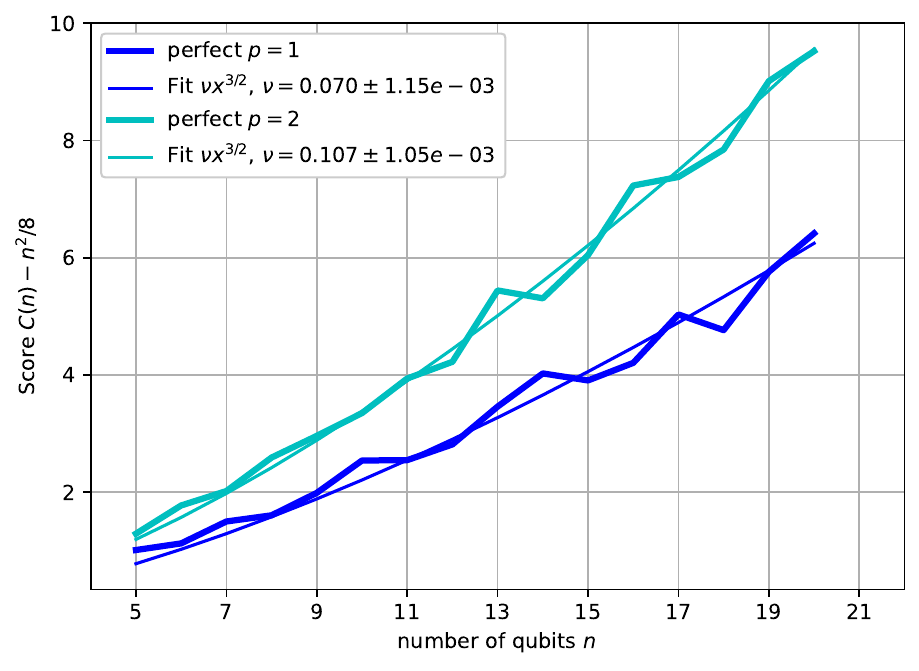}  \\

    \caption{\emph{Top}: Scaling of the expected maximum cut size for Erd\"os-Renyi graphs of increasing size. Each data point (in blue) is computed by solving $200$ MaxCut instances (using the AKMaxSAT solver \cite{10.1007/11499107_27}). In orange is a fit of shape $y=\frac{n^2}{8} + \lambda n^{\frac{3}{2}}$ with $\lambda\approx 0.178$ obtained by a standard least-squares method (r-value $> 1 - 10^{-3}$). \emph{Bottom}: Fit of QAOA scores $C(n) - n^2/8$ to $\nu n^{3/2}$ for $p=1$ (blue) and $p=2$ (cyan). The obtained values for $\nu$ correspond to $\beta = \nu / \lambda = 40\%$ and $60\%$, respectively.}
    \label{fig:maxcut_scaling}
\end{figure}

Since we are interested in a typical behavior over a class of random graphs, we want to average this quantity, giving us an expected approximation ratio over instances of a given size, $$\overline{\alpha}^\mathrm{Q}(n)=\mathbb{E}_{G \sim \mathcal{G}(n, \frac{1}{2})} \left[\alpha^\mathrm{Q}(G)\right].$$ The behavior of this quantity is hard to derive, but it is easy to derive the behavior of the closely related quantity, $$\alpha^\mathrm{Q}(n) \equiv \frac{\mathbb{E}_{G \sim \mathcal{G}(n, \frac{1}{2})}\left[\mathbb{E}^\mathrm{Q}[C(G)]\right]}{\mathbb{E}_{G \sim \mathcal{G}(n, \frac{1}{2})}[C_\mathrm{max}(G)]}\equiv\frac{C^\mathrm{Q}(n)}{C_\mathrm{max}(n)}.$$
$\alpha^\mathrm{Q}(n)$ can be seen as a first-order approximation to $\overline{\alpha}^\mathrm{Q}(n)$.
Since QAOA produces score distributions that are \emph{at least} as good as straightforward random sampling, we get 
\begin{equation}
  \alpha^\mathrm{Q}(n)\geq \frac{C^\mathrm{R}(n)}{C_\mathrm{max}(n)}.
  \label{eq:inequality}
\end{equation}
We now turn to the behavior of $C^\mathrm{R}(n)$ and $C_\mathrm{max}(n)$.
Erd\"os-Renyi graphs of $\mathcal{G}\left (n, \frac{1}{2}\right)$ have, on average, $\frac{n^2}{4}$ edges. On average over the complete family, their cuts have an expected cost $$C^\mathrm{R}(n)\equiv\mathbb{E}_{G \sim \mathcal{G}(n, \frac{1}{2})}\left[\mathbb{E}^\mathrm{R}[C(G)]\right] = \frac{\mathbb{E}[|E|]}{2} = \frac{n^2}{8}.$$ Recent results \cite{gamarnik2014,dembo2017} show that their typical maximum cut size grows as 
\begin{equation}
    C_\mathrm{max}(n)\equiv\mathbb{E}_{G \sim \mathcal{G}(n, \frac{1}{2})}(C_\mathrm{max}(G))=\frac{n^2}{8} + \lambda n^{\frac{3}{2}} +o(n^{3/2}),
    \label{eq:Cmax}
\end{equation}
with $\lambda \geq \frac{1}{2\sqrt{\pi}}\approx 0.159$. In practice, a numerical fit in the range $n\in[5, 40]$ yields a value of $\lambda \approx 0.178$ (see Figure \ref{fig:maxcut_scaling}).
Plugging these results into Eq.~(\ref{eq:inequality}), we obtain
\begin{equation}
  \alpha^\mathrm{Q}(n)\geq  \frac{\frac{n^2}{8}}{\frac{n^2}{8} + \lambda n^{\frac{3}{2}}}, 
  \label{eq:inequality_full}
\end{equation}
which approaches $1$ when $n$ diverges.

\paragraph{An improved approximation ratio}
This lower bound suggests that $\alpha^\mathrm{Q}(n)$ is not the appropriate quantity to consider to assess the quality of a heuristic for MaxCut on this class of graphs.
Because the expected approximation ratio of random sampling grows with $n$, requiring a quantum processor to achieve a fixed approximation ratio is not an interesting test (for this class of graphs): this ratio will get easier and easier to reach as $n$ grows. 
For instance, the previous inequality tells us that over random graphs of size $1500$, random sampling will produce cuts with an average score that is $99.5\%$ of the average score of the maximal cuts.
This means that the most ineffective quantum processor, as long as it has $1500$ qubits, will achieve at least the same ratio of expected cost. This phenomenon was for instance observed in \cite{dalyac2020qualifying}, where both random and quantum approaches seemed to behave increasingly well for larger instances. 
This behavior is not a particularity of the $\mathcal{G}(n, \frac{1}{2})$ class. In fact, this result holds for any class of random graphs such that edges are picked uniformly at random \cite{gamarnik2014,dembo2017} and such that the number of edges grows faster than $O(n)$. If the number of edges is a $O(n)$, then the standard average approximation ratio definition will be upper bounded by a constant that can be analytically derived. This is for instance the case for $k$-regular graphs (see section \ref{ssec:graphs}). 

In order to avoid this issue, we consider instead the same quantities after subtracting the leading $\frac{n^2}{8}$ term:
\begin{equation}
    \beta(n) \equiv\frac{C(n)-\frac{n^2}{8}}{C_\mathrm{max}(n)-\frac{n^2}{8}}=\frac{C(n)-\frac{n^2}{8}}{\lambda n^{3/2}},
    \label{eq:qscore_def_detailed}
\end{equation}

We use this definition to specify the conditions to pass the Q-score: we require the quantum algorithm achieve a ratio that exceeds a constant value $\beta^\star \in ]0, 1[$:
\begin{equation}
    \beta^\mathrm{Q}(n) \geq \beta^\star.    
\end{equation}
Based on numerical simulations with NISQ-compatible noise levels (see subsection \ref{subsec:illustration} above), we fix $\beta^\star$ to $\beta^\star = 20\%$.

This requirement implies that the quantum heuristic must fulfill satisfactory scalability properties:
indeed, achieving a ratio $\beta^\mathrm{Q}_n \geq \beta^\star$ implies that the quantity $C^\mathrm{Q}(n) - \frac{n^2}{8}$ grows at least as $\nu n ^ {3/2}$, with $\nu=\beta^\star\lambda$ and $\lambda$ the scaling of the optimal solution (see Eq.~\eqref{eq:Cmax}). 
In other words, we require the scaling rate of the quantum heuristic to be at least within a fraction $\beta^\star=20\%$ of the scaling of the optimal solution.

For instance, random sampling, which always produces a vanishing ratio $\beta^\mathrm{R}(n)=0$, cannot fulfill the Q-score for any $\beta^\star>0$.
Conversely, requiring $\beta^\star = 100\%$ would mean requiring to achieve the optimal solution. 

\begin{figure*}
    \centering
         (a)\includegraphics[width=0.85 \columnwidth]{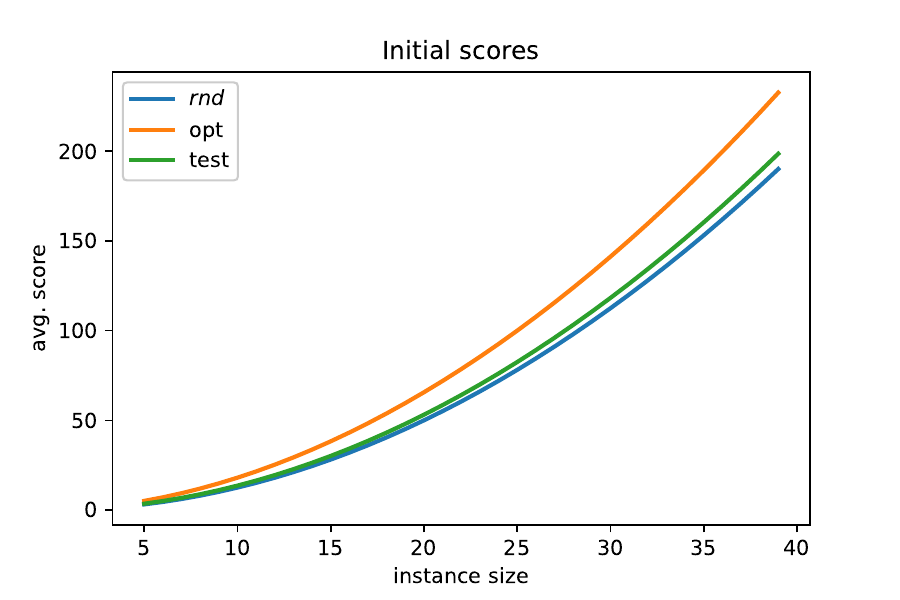}
         (b)\includegraphics[width=0.85 \columnwidth]{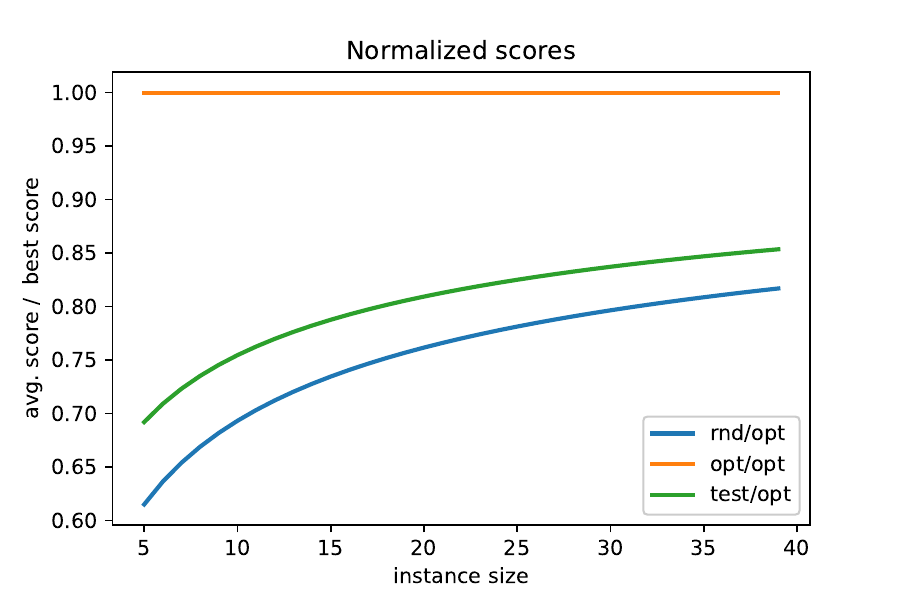}  \\
         (c)\includegraphics[width=0.85 \columnwidth]{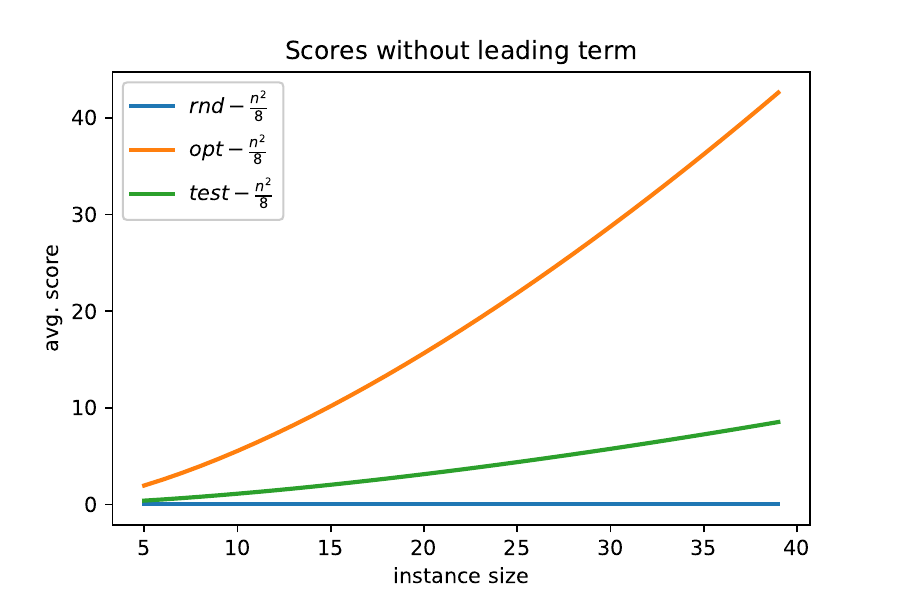}
         (d)\includegraphics[width=0.85 \columnwidth]{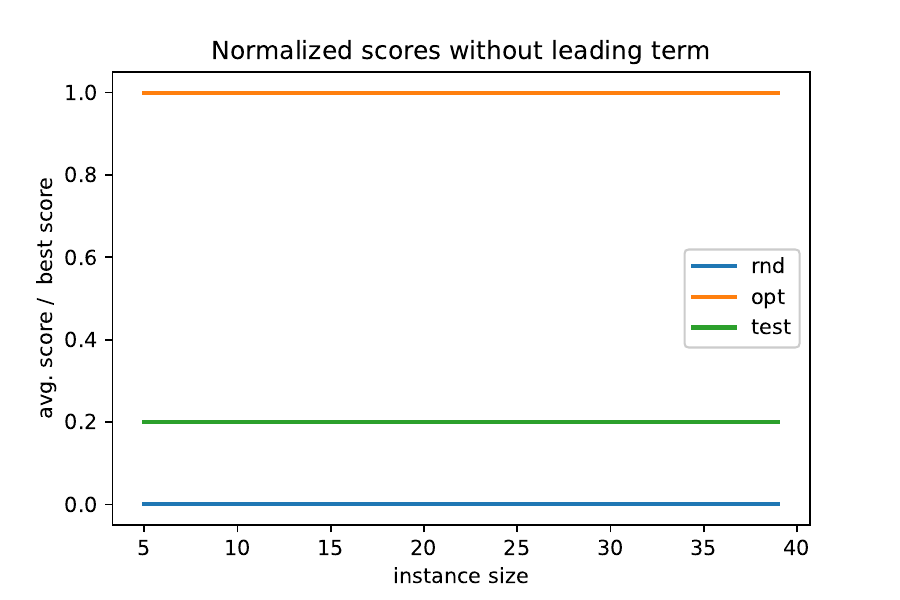}  \\
    (e)\includegraphics[width=0.75 \columnwidth]{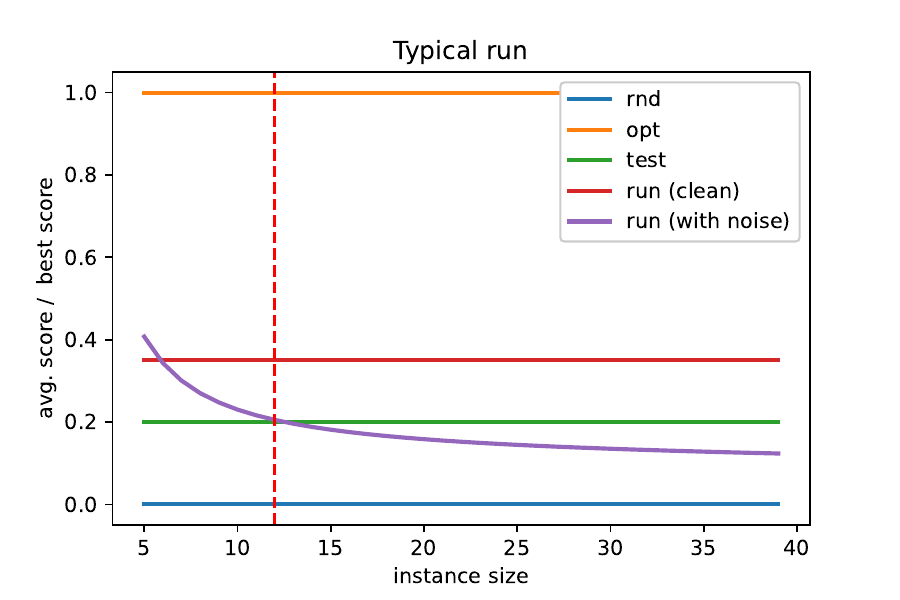}
    \caption{(a) Typical scaling of the expected costs $C(n)$ for three cases: expected maximum cut size $C_\mathrm{max}(n)$ (orange), expected random cut size $C^\mathrm{R}(n)$ (blue), and cost corresponding to our threshold $\beta^\star = 20\%$ (green). (b) Scaling of the average expectation ratios $\alpha(n)$ (namely cost normalized by the expected maximum cut size $C_\mathrm{max}(n)$) (c) Scaling of the cost with the leading $n^2/8$ term  subtracted. (d) Scaling of the improved expectation ratios $\beta(n)$ (namely with the leading term subtracted and normalized by the maximum cut scaling). (e) Evolution of $\beta(n)$ for a typical Q-score run: in red, the scaling of a QAOA running on a perfect quantum processor. In purple, the scaling of a QAOA running on a imperfect processor. In this last setting, the dashed red line will give the returned Q-score.}
    \label{fig:recap_metric}
\end{figure*}

Figure \ref{fig:recap_metric} gives a qualitative graphical summary of the different quantities discussed here.

\paragraph{Remark.} In \cite{arute2020quantum}, the authors use a similar definition for the approximation ratio, with different motivations. Their definition comes from the fact that they are interested in minimizing the energy of Ising Hamiltonians of shape $ H_G' = \sum_{i, j\in E} \sigma_z^{(i)}\sigma_z^{(j)}$, i.e. without the constant energy offset of $\frac{|E|}{2}$ (compare to $H_G$ in Eq.~\eqref{eq:HG_def}). The spectra of these Hamiltonians do not coincide with the usual cut size functions, but exhibit the same feature as the cost metric described in Eq.~(\ref{eq:qscore_def_detailed}).

\paragraph{A continuous score.} Even though the proposed protocol outputs a single number, it is possible to extract far more information from a run of Q-score.
For instance, a good benchmark metric would be to track the largest $\nu$ constant accessible for each problem size $n$. This scaling would allow a manufacturer to track the performances of its processors when scaling up the number of qubits/problem size. Moreover, this $\nu$ factor provides a comparison tool with various behaviors whether it is random sampling ($\nu = 0$), perfect solving ($\nu = \lambda \approx 0.178$), perfect QAOA ($\nu \approx 0.07$ for $p=1$, $\approx 0.107$ for $p=2$, see Fig.~\ref{fig:maxcut_scaling}).

\subsection{A note on the experimental parameters}

When defining the protocol, we set the value of the number of shots as well as the optimization procedure.
While these choices are somewhat arbitrary, they arguably do not significantly impact the final value of the Q-score.

The number of shots (2048) is representative of the typical numbers of shots used on experimental processors. It gives reasonable statistical errors on the estimate of the cost function. Given the wide range of clock speeds of existing QPUs, this fixed number of shots does not yield equivalent run times. These clock speeds, or equivalently,  the time-to-solution, could easily be taken into account by Q-score by setting a maximum time budget to compute $\beta^\mathrm{Q}(n)$ for a given graph size $n$.
For the time being, we did not specify such a time limit, since in the current state of QPUs, increasing the processor’s fidelities probably comes before increasing their speed. But Q-score should be reported together with the absolute time required to compute $\beta(n^\star)$. 

As for the choice of the classical optimization procedure: 
%we argue that the value of Q-score does not depend on the classical optimizer, provided it is "good enough", i.e it suitably optimizes the QAOA parameters. In our experience, COBYLA is one such optimizer.
here, we took the COBYLA optimizer that is very commonplace and usually provides a good tradeoff between convergence speed and quality of the attained minimum \cite{Lavrijsen2020,Khairy2019} (although, as a local optimizer, it may have issues in the presence of flat surfaces---issues which we did not observe for the parameter ranges we investigated). Beyond this particular choice, we stress that the same optimizer should be used across all platforms in order to ensure a consistency in the obtained scores.

\subsection{Changing the graph class} \label{ssec:graphs}
 In this protocol, and the discussion of section \ref{subsec:ratio}, we focused on a particular class of random graphs, namely Erd\"os-Renyi random graphs with edge probability $\frac{1}{2}$. These graphs have the nice property of being dense, and thus any (positive) result for this class of graph has a good chance to transpose to any application.
 However, running QAOA-MaxCut for these graphs can be quite demanding, since a typical circuit would have around $k\frac{n^2}{2}$ CNOT gates for an Ansatz of depth $k$ over a graph of size $n$. This quadratic scaling can be quite demanding for a real hardware platform.
 
 In this section, we show how a similar score/test can be derived for other classes of random graphs that would define less demanding tests, as in running circuits with a lower entangling gate count. All the results presented below can be derived from the scaling proven in \cite{dembo2017}. In this work, the authors state that the scaling of the average maximum cut size for random graphs with $\gamma n$ edges picked uniformly can be expressed as:
\begin{align}
    C_{\textrm{max}}(n) = \frac{n\gamma}{2} + P_\star \sqrt{\frac{\gamma}{2}}n + o(n\sqrt{\gamma}) \label{eq:maxcutscaling}
\end{align}
 where $P_\star \leq \sqrt{2/\pi}$. Numerical estimate of this constant gives $P_\star = 0.76321 \pm 0.00003$.
 This result gives us quite naturally the difference in scaling between the cut sizes produced by random sampling, $\frac{n\gamma}{2}$, and the cut sizes produced by an exact solver.
 
 We now detail this scaling for two classes of graphs: generic $\mathcal{G}(n, p)$ random graphs and random $k$-regular graphs.
 
 \paragraph{$\mathcal{G}(n, p)$ graphs} We can run the same calculation as the one for $\mathcal{G}(n, \frac{1}{2})$ for any edge probability $p$. In this setting, we have $\gamma = \frac{pn}{2}$ and similarly to the $\mathcal{G}(n, \frac{1}{2})$ case, the average maximum cut size grows as $$C_{\textrm{max}}(n) = p\frac{n^2}{4} + \lambda_p n^{\frac{3}{2}} $$
for some constant $\lambda_p$. Analytically, we expect $\lambda_p = \sqrt{2p}\lambda_{\frac{1}{2}}$, with $\lambda_{\frac{1}{2}}$ the scaling of the $p = \frac{1}{2}$ case. The direct consequence is that we can use a similar test as for $p=\frac{1}{2}$ and pose:
$$\beta(n) = \frac{C(n) - p\frac{n^2}{4}}{C_{\textrm{max}}(n) - p\frac{n^2}{4}} = \frac{C(n) - p\frac{n^2}{4}}{\lambda_p n^{3/2}}$$
where $\lambda_p$ can be either fitted numerically or taken as $\lambda_{\frac{1}{2}}\sqrt{2p}$. Overall, this boils down to comparing the QAOA performance $C(n) - p \frac{n^2}{4}$ against a $n^{3/2}$ scaling.

Here, we derived an expression for $\beta(n)$ where $p$ is constant, but the derivation hold for any size dependent probability $p=f(n)$. Hence, we can define the same benchmark with increasingly dense (and thus difficult to implement) instances.
 
\paragraph{$k$-regular graphs} Regular graphs have the convenient property of being very sparse, with a number of edges of $\frac{kn}{2}$ for a $k$-regular graph of size $n$. For this class of graph the scaling of the average maximum cut is in fact proven, and not only known within an interval.
Applying Eq.~(\ref{eq:maxcutscaling}) with $\gamma = k/2$ gives us:

$$ C_\mathrm{max}(n) = \frac{nk}{4} +\frac{P_\star\sqrt{k}}{2}n + o(n\sqrt{k}),$$
hence a natural choice of $\beta$ is:

$$\beta(n) = \frac{C(n) - \frac{nk}{4}}{C_{\textrm{max}}(n) - \frac{nk}{4}} = \frac{C(n) - \frac{nk}{4}}{\lambda n}$$
for some constant $\lambda = P_\star\sqrt{k}/2$. Once again, we can either use the analytical value for $\lambda$ or fit it numerically for small instances.
That is, if we fix $k$, we are looking to compare the QAOA performances over $k$-regular graphs $C(n) - \frac{nk}{4}$, to a linear scaling in $n$.

\section{Running Q-score yourself: an open-source repository}\label{sec:doi}

We provide a Python package, \verb+qscore+ (\url{https://www.github.com/myQLM/qscore}), to compute the Q-score for any QPU that has been interfaced with the open-source \verb+myqlm+ library.
Once the \verb+qscore+ package is installed, here is the typical script that needs to be run:

\begin{lstlisting}[language=Python, caption=Python script to run Q-score]
from qat.qscore.benchmark import QScore
from qat.plugins import ScipyMinimizePlugin
from qat.qpus import get_default_qpu

# Our QPU is composed of:
# - a variational optimizer plugin
# - a QLM/myQLM default qpu (either LinAlg or pyLinalg)

QPU = ScipyMinimizePlugin(
    method="COBYLA", 
    tol=1e-4, 
    options={"maxiter": 300}
) | get_default_qpu()

benchmark = QScore(
    QPU,
    size_limit=20,  # limiting the instace sizes to 20
    depth=1,        # using an Ansatz depth of 1
    output="perfect.csv",
    rawdata="perfect.raw"
)
benchmark.run()
\end{lstlisting}

Here, the QPU is a perfect circuit simulator provided by myQLM. In order to use a true hardware QPU, one simply needs to interface
 one's QPU with the myQLM API. This thin layer typically looks as follows:
 \begin{lstlisting}[language=Python, caption=Python script to make your own QPU compatible with myQLM]
from qat.core.qpu import QPUHandler
from qat.core import Result


class MyQPU(QPUHandler):
    def submit_job(self, job):
        # Evaluate the job using your QPU
        # A job constains:
        # a circuit:
        circuit = job.circuit
        # possibly an observable
        observable = job.observable
        # or a list of qubits to sample:
        qubits = job.qubits

        # Results are returned in a `Result` object
        return result
 \end{lstlisting}
 
\section{Conclusion}

In this note, we have introduced the Atos Q-score, an application-centric, hardware-agnostic and scalable metric that measures the ability of a full quantum stack---hardware and software---to solve a prototypical combinatorial optimization problem, MaxCut, using the Quantum Approximate Optimization Algorithm, a widespread variational quantum heuristic compatible with Noisy Intermediate Scale Quantum co-processors.
Instead of focusing on how well the basic building blocks of a quantum processor work, like most existing metrics, Q-score provides information as to the capacity of the processor to solve an actual problem. It does so without favoring any hardware technology or software paradigm, and will be applicable to very large problems due to its scalability.

Like the classical LINPACK benchmark, the Q-score focuses on a given problem as a proxy for most other hard computational problems. Here, MaxCut was chosen as a representative hard problem, because it appears to be quite simple and universal. In the search for the "killer application" for quantum co-processors, other more relevant problems may appear and supersede MaxCut, but the same strategy as the one we describe in this note will likely be applicable.
Likewise, the choices of optimizer (COBYLA) and other parameters (number of shots, number of graphs, etc) we set the value of for the sake of standardization have a degree of arbitrariness.
In a similar vein, the current protocol is geared to digital quantum co-processors. An extension to analog processors is rather straightforward, and will be the topic of future work.

All these variations on the protocol proposed in this note should not influence the overall outcome of the procedure, and thus the usefulness of the benchmark.

\section*{Acknowledgements}
We acknowledge useful discussions with the members of the Atos Quantum Advisory Board, Alain Aspect, David DiVincenzo, Artur Ekert, Daniel Est\`eve, and Serge Haroche. The computations have been performed on the Atos Quantum Learning Machine.

\bibliographystyle{alpha}
\bibliography{biblio.bib}
\EOD
\end{document}